\documentclass[9pt,twocolumn,twoside]{opticajnl}
\journal{opticajournal} 

\setboolean{shortarticle}{true}

\usepackage{makecell}
\usepackage{xcolor}
\usepackage{hyperref}


\usepackage{color}

\title{Vertically-coupled microring resonators for lithium niobate on insulator integrated photonics}

\author[1]{Clément Selva}
\author[1]{Sarah Umme Tabassum}
\author[1]{Kumar Kinjalk}
\author[1]{Adrian Laborde}
\author[1]{Antoine Monmayrant}
\author[1]{Olivier Gauthier-Lafaye}
\author[1]{Stéphane Calvez}

\affil[1]{LAAS-CNRS, Université de Toulouse, CNRS, Toulouse, France}

\affil[*]{stephane.calvez@laas.fr}

\begin{abstract}

We propose and demonstrate lithium niobate on insulator (LNOI) microring resonators that are vertically-coupled to index-matched silicon-rich silicon nitride (SiN$_x$) access waveguides as an alternative to conventional lateral coupling schemes for integrated photonics applications. The wafer-scalable fabrication process involves a planarization step exploiting a single-step and reproducible spin-coating of a SU-8 epoxy spacer layer which allows a controlled separation to be established between the resonator and its access waveguide. The resulting multimode and single-mode microring resonators exhibit, for their fundamental TE$_0$ mode, intrinsic quality factors of $\mathbf{4.0\times10^{5}}$ and $\mathbf{3.4\times10^{5}}$, respectively.

\end{abstract}

\setboolean{displaycopyright}{false} 

\begin{document}

\maketitle

\section{Introduction}

Over the past decade, lithium niobate on insulator (LNOI) has emerged as a leading platform for nonlinear integrated photonics for the implementation of both classical \cite{Wang2019} and quantum applications \cite{Moody2022}, owing to the ability to fabricate tightly-confined optical waveguide structures with a material presenting favorable properties including a broad transparency window extending from $0.4$ to 5 µm \cite{Boes2023}, a low level of optical loss \cite{Younesi2025}, and large second-order ($\chi^{(2)}$) \cite{Nikogosyan2005} and third-order ($\chi^{(3)}$) \cite{Schiek2025} nonlinear susceptibilities. 

Instrumental to the demonstration of compact yet efficient nonlinear devices is the use of Whispering Gallery Mode resonators such as microring or microdisk resonators which are resonant structures with enhanced intracavity fields, allowing efficient light--matter interactions. On the LNOI platform, microring resonators have, more specifically, enabled efficient Second-Harmonic Generation (SHG) \cite{Cheng2023,Wu2024} and Spontaneous Parametric Down-Conversion \cite{Chen2025}, both relying on $\chi^{(2)}$ nonlinear interactions, as well as Four-Wave Mixing \cite{Wang2019} based on the $\chi^{(3)}$ nonlinear interactions. In all cases, the resonators are laterally coupled to their access waveguide(s) since this coupling scheme benefits from a fabrication involving only one lithographic/etching step. The downside is the requirement to meet demanding constraints in terms of the waveguide lateral profiles (needing to present as vertical sidewalls as possible) and a tight control of the separation. 
A contrario, the vertical coupling architecture, where the access waveguides lie above or below the resonator, offers an attractive alternative by replacing the lateral gap with a vertically controlled spacer layer, thereby providing an additional degree of freedom to tailor the coupling as both the vertical separation and the lateral offset between the ring and the access waveguide can serve as independent adjustment parameters. Furthermore, this approach facilitates the heterogeneous integration between different material platforms as demonstrated, for instance, for the integration of III-V semiconductors or silicon nitride (SiN) with silicon (Si). The critical technological challenge is the realization, over a patterned structure, of a planarized spacer layer with precise thickness control. 
Several strategies have been explored to make such vertically-coupled photonic structures on the various platforms of relevance, exploiting either the localized oxidation of the spacer layer \cite{Koonath2004,Calvez2015}, the deposition and subsequent polishing of dielectric spacers \cite{Feng2014}, or the spin-coating in one or multiple steps of polymeric \cite{Ghulinyan2011} \cite{Absil2001} or spin-on-glass spacers \cite{Kokubun2005}.

In this Letter, we demonstrate vertically-coupled LNOI microring resonators with index-matched SiN$_x$ access waveguides whose integration relies on a planarizing SU-8 vertical spacer layer realized using a simple yet reproducible spin coating and a controlled slow-ramp bake. This technological approach benefits from being wafer-scalable contrary to the focused ion beam fabrication scheme reported in reference \cite{zhang_MDPI2021}. Furthermore, this demonstration extends the recent reported work on heterogeneous integration of lithium niobate (LN) devices which has so far been focused either on fabricating SiN-strip-loaded LN waveguides using micro-transfert printing \cite{Niels2026,Sumon2026} or on obtaining vertical coupling between BCB-bonded SiN and LN ridge injection waveguides for electro-optic modulator integration \cite{Ruan2023}.

\smallskip

\section{Device design}
\label{sec:vertical_structure_design}

\begin{figure}[!ht]
	\centering
	\includegraphics[width=\linewidth]{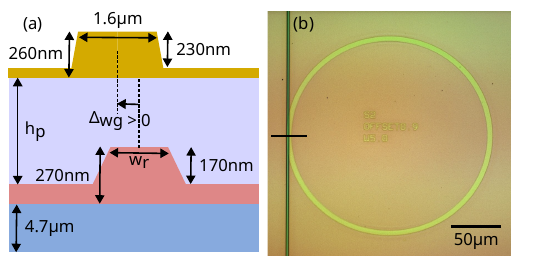}
	\caption{Geometry of the vertical structure. (a) Cross-section of the coupling region showing in blue the SiO$_2$ buried oxide (BOX),
   in red the LN rib ring resonator with top width $w_r$, in purple the SU-8 spacer of thickness $h_p$, and in gold the SiN$_x$ access waveguide laterally shifted by $\Delta_{\mathrm{wg}}$ ( $\Delta_{\mathrm{wg}}>0$ corresponding to an outward shift of the access waveguide relative to the ring center). The LN sidewalls are angled by $25^\circ$ \cite{Kaufmann2023}. (b) Top-view confocal microscope image of a fabricated $R=100$\,µm resonator with a ring width $w_r=5$\,µm, overlayed with a $h_p=1.52$\,µm SU-8 spacer layer and an access waveguide offset by $\Delta_{\mathrm{wg}}=0.9$\,µm. The black line depicts where cross-section (a) was taken.}
	\label{fig:vertical_coupling_structure}
\end{figure}

The considered vertical architecture consists of a LNOI microring resonator located below a SiN$_x$ single-mode (TE$_0$) access waveguide, separated by a planarized SU-8 spacer, as illustrated in Fig.~\ref{fig:vertical_coupling_structure}. Both single transverse mode (of width $w_r = 1.1$ µm) and multimode  ($w_r = 5$ µm) rings are considered hereafter.

The LN ring resonators are partially etched, leaving a $100\,\mathrm{nm}$-thick slab beneath the rib as depicted in Fig.~\ref{fig:vertical_coupling_structure} (a). This geometry was selected as the optimal trade-off between reducing scattering losses induced by sidewall roughness and maintaining tight optical confinement \cite{Selva2026}. A ring radius of 100 µm was chosen as a compromise between compactness and negligible bending losses.

The access waveguide material was chosen to be SiN$_x$ as it is a widely established material for photonics whose processing conditions can be selected to enable precise control of film thickness, stoichiometry, and consequently refractive index.
The SiN$_x$ waveguide dimensions (Fig.~\ref{fig:vertical_coupling_structure} (a)) are chosen to achieve effective index matching with the fundamental TE$_0$ mode of the single-mode LN ring resonator, both being calculated using the FEMWELL open-source mode solver \cite{Gehring}. For the multimode geometry, the effective indices of the TE$_0$, TE$_1$, and TE$_2$ modes remain sufficiently close to that of the SiN$_x$ waveguide to ensure a beat length larger than the interaction length, enabling efficient coupling to all three modes.

Once the waveguide geometries are fixed, the propagation losses of the LN ring are calculated using a roughness-induced scattering model based on \cite{Selva2026}, leading to an estimated intrinsic quality factor at a wavelength of 1550\,nm of 3.2$\times10^{5}$ for the singlemode microring and a corresponding nominal critical coupling coefficient of $\approx$0.025. 

\begin{figure}[!ht]
	\centering
	\includegraphics[width=\linewidth]{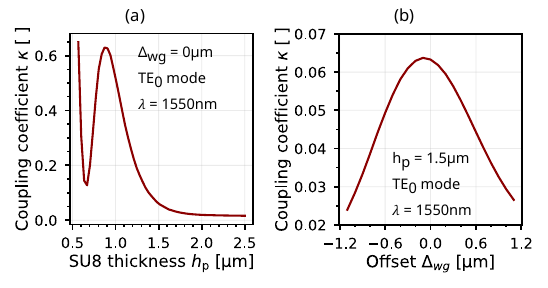}
	\caption{Simulated SiN$_x$--LN power transfer for a ring width of $w_r=1.1$ µm. (a) For different SU-8 spacer thickness $h_p$. (b) For different lateral offset $\Delta_{\mathrm{wg}}$.}
	\label{fig:coupling_transfer}
\end{figure}

The latter coupling coefficient between the SiN$_x$ access waveguide and the LN ring resonator is ruled by their relative transverse positions and overlapping interaction length. The related calculations are performed using the eigenmode expansion (EME) solver in Lumerical \cite{Optics}. As alluded to in the introduction, the coupling can be tailored by varying the vertical separation, $h_p$, and the lateral offset, $\Delta_{\mathrm{wg}}$, seeking suitable positions allowing the critical coupling condition to be fulfilled. Although a full bi-dimensional mapping could be carried out and reported, only single parameter evolutions are presented hereafter (in Fig.~\ref{fig:coupling_transfer}) as they allow a full description of the physics at play. First, the influence of the SU-8 spacer thickness $h_p$ is investigated (Fig.~\ref{fig:coupling_transfer} (a)). Reducing the spacer thickness increases the evanescent field overlap between the two waveguides, resulting in stronger coupling. For SU-8 thicknesses below $1\,\mu$m, the coupling coefficient first approaches unity and then decreases due to the periodic power exchange between the two waveguides.  Indeed, as the modal overlap increases, the coupling length becomes shorter than the effective interaction length, resulting in power being transferred back to the access waveguide \cite{Ghulinyan2013}. A spacer thickness of $h_p=1.5$\,µm is selected to provide an over-coupled initial condition. The transferred power can be fine-tuned by varying the lateral offset $\Delta_{\mathrm{wg}}$ to reach the critical coupling condition.

Starting from this overcoupled situation, the lateral SiN$_x$/LN alignment can then be optimized by varying the lateral offset around the centered configuration ($\Delta_{\mathrm{wg}}=0$\,µm), with $\Delta_{\mathrm{wg}}\in[-1.1,1.1]$\,µm. This radial displacement induces simultaneously a change in the modal overlap and the effective interaction length between the resonating and access waveguide modes. As shown on Fig.~\ref{fig:coupling_transfer} (b), for large offsets, the decrease in modal overlap is the dominant factor and the coupling coefficient becomes weak. In the single transverse mode resonator situation, near the centred configuration, the coupling is mainly ruled by the interaction length and the maximum coupling coefficient is achieved for an access waveguide positioned slightly inwards ($\Delta_{\mathrm{wg}}=-0.1$\,µm).

\section{Device fabrication}

The devices were fabricated on a $16\times10\,\mathrm{mm}^2$ sample diced from a 4-inch LNOI wafer supplied by InnoSemi. The wafer consists of an X-cut LN thin film ($\sim300\,\mathrm{nm}$ thick) bonded to a 4.7\,µm-thick plasma-enhanced chemical vapor deposition (PECVD) SiO$_2$ BOX layer on a 525\,µm-thick Si substrate. The sample was cleaned in acetone (ACE) and isopropanol (IPA) for 5\, min, followed by an oxygen plasma treatment to remove residual organic contaminants.
Three identical sets of ten devices were fabricated for each microring rib width ($w_r=1.1$ and $w_r=5$\, µm). One of the fabricated devices is shown in Fig.~\ref{fig:vertical_coupling_structure} (a). For both geometries, the lateral offset was varied over $\Delta_{\mathrm{wg}}\in[-0.7,1.1]$\,µm with a step size of 0.2\,µm.

 The LN ring resonators were then patterned by Electron Beam Lithography (EBL, Raith 150) using a 400-nm-thick ma-N 2405 negative-tone resist. To mitigate charging effects during exposure, the resist was coated with Electra 92 conductive polymer. The EBL exposure was performed at an acceleration voltage of 20\,kV with a nominal dose of 130\,µC/cm$^2$. After exposure, the conductive layer was removed in deionized water for 1\,min, and the resist was developed in MF-26-CD developer for 45\,s, followed by a 1\,min deionized water rinse. The resulting resist pattern was then transferred into the LN layer by inductively coupled plasma reactive ion etching (ICP-RIE, Sentech SI500F) using a pure argon physical etching process adapted from \cite{Kaufmann2023} to limit the residual byproducts. A subsequent ACE and IPA bath dipping and oxygen-plasma stripping step was performed to remove the remaining resist and etching residues. Finally, the LN surface was cleaned using a Standard Clean 1 (SC-1) process for 30\,min at $65^\circ$C following the procedure reported in \cite{Kinjalk2025}, resulting in a final LN film thickness of 270\,nm and a remaining slab thickness of 100\,nm. The fabricated ring resonators feature top rib widths of 1.1\,µm and 5\,µm.

Following LN ring fabrication, the vertical spacer layer was defined by spin-coating SU-8 6002.2, a negative-tone epoxy photoresist. The resist planarization involved a reflowing step performed at $110\,^\circ\mathrm{C}$ using $4\,^\circ\mathrm{C/min}$ ramps, and a 3-min-long $150\,^\circ\mathrm{C}$ hardning step. 

A $260\,\mathrm{nm}$-thick SiN$_x$ film was subsequently deposited at $100\,^\circ\mathrm{C}$ by inductively-coupled plasma-enhanced chemical vapor deposition, yielding a refractive index of 2.23 at 1550\,nm. The SiN$_x$ access waveguides and the chip in/output grating couplers (GCs) with tapers were then fabricated using the same lithography and etching process as the LN ring resonators, omitting the final SC-1 cleaning step.

\begin{figure}[!ht]
	\centering
	\includegraphics[width=\linewidth]{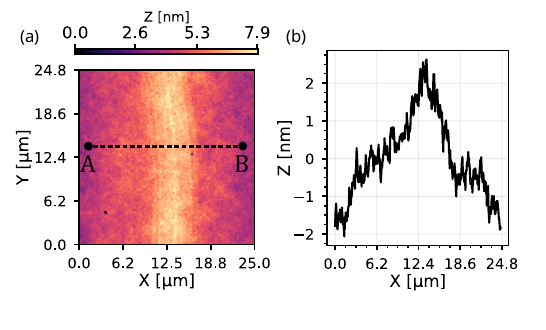}
	\caption{Atomic Force Microscopy measurement of the planarized surface topography of the device shown in Fig.~\ref{fig:vertical_coupling_structure}\,(a))
	(a) Spatial distribution and (b) Quantitative height profile extracted along the A--B cross section indicated in (a).}
	\label{fig:AFM}
\end{figure}

To assess the quality of the planarization, reflectometry mapping of the whole sample was performed using, on the illumination side, a white LED whose beam is focused down to a 22-µm-waist spot onto the sample using a 10x NA=0.3 objective and, on the detection side, a Si-based spectrometer system with a spectral resolution of 0.95\,nm. In the centre of the sample, the SU-8 spacer thickness was found to be of $1.52\pm0.03$\,µm. The SU8 thickness was observed to vary by less than 103\,nm (corresponding to the observation quarterwavelength) over the surface that is unaffected by the $\sim$3.5-mm-long edge beads induced by the relatively slow spinning speed used. To complement the latter wide-area characterization, atomic force microscope (AFM) measurements were performed after SiN$_x$ waveguide fabrication over a $25\times24.8$\,µm$^2$ area located on the remaining SiN$_x$ slab above a 5-µm-wide ring resonator, as shown in Fig.~\ref{fig:vertical_coupling_structure} (a). The measured topography, after tip/tilt plane removal, exhibits a peak-to-valley height variation of $S_z = 7.92\,\mathrm{nm}$. The residual height modulation observed along the A--B cross section in Fig.~\ref{fig:AFM} (a) originates from the buried ring resonator and remains limited to a few nanometers. These results underline the effectiveness of the planarization process in preserving a quasi-flat surface, even above the buried LN structures.

\section{Device characterization}

The transmission measurements were performed using a tunable laser (Santec TSL-570), providing a wavelength sweep from 1500 to 1620\,nm. The laser output was first passed through a polarization controller to optimize the input polarization before being coupled into the device through the input GCs. The transmitted light was collected from the output grating coupler and detected using an InGaAs amplified detector (PDA20CS-EC).

The measured resonances exhibit Fano-like line shapes (see Fig. \ref{fig:single_mode_transmission} and \ref{fig:multimode_transmission}) because the resonators are excited with a slightly bimodal injection waveguide \cite{Ding14} due to the presence of the marginally-too-thick residual LN slab. 

\subsection{Single-mode microring resonator}

\begin{figure}[!ht]
	\centering
	\includegraphics[width=\linewidth]{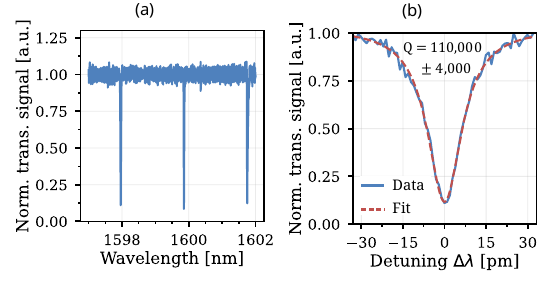}
	\caption{Transmission characterization of a single-mode LN ring resonator with $w_r=1.1\,\mu\mathrm{m}$ and $\Delta_{\mathrm{wg}}=0.9\,\mu\mathrm{m}$. (a) Normalized transmission spectrum over two FSRs. (b) Resonance near $\lambda=1598\,\mathrm{nm}$ with symmetrical Fano fit.}
	\label{fig:single_mode_transmission}
\end{figure}

Fig.~\ref{fig:single_mode_transmission} presents the optical response of a single-mode LN microring resonator with a rib width of $w_r=1.1\,\mu\mathrm{m}$ and a positive lateral offset $\Delta_{\mathrm{wg}}=0.9\,\mu\mathrm{m}$, selected on the basis that the device operates in a regime close to the critical coupling condition. The normalized transmission spectrum, recorded over two free spectral ranges (FSRs) between 1597 and 1601 nm, exhibits regularly spaced resonance dips corresponding to the coupling with the fundamental TE$_0$ mode. The measured FSR of $\sim1.92\,\mathrm{nm}$ yields a group index of $n_g\sim2.12$, in agreement with the expected value.

An enlarged view of the resonance around $\lambda\approx1598\,\mathrm{nm}$ is shown in Fig.~\ref{fig:single_mode_transmission} (b). A Fano-shape fit \cite{Ding14} gives a resonance linewidth of $\Gamma\approx15\,\mathrm{pm}$, corresponding to a loaded quality factor of $Q_\mathrm{L}=1.1\times10^{5}$ with a 95\% confidence interval of $\pm 4\times10^{3}$. Considering the slightly over-coupled regime predicted by simulations in~\ref{sec:vertical_structure_design} and the measured transmission minimum of $T_\mathrm{min}=0.12$, the intrinsic quality factor is estimated as $Q_\mathrm{i}=2Q_\mathrm{L}/(1-\sqrt{T_\mathrm{min}})\approx3.4\times10^{5}$.

\subsection{Multimode microring resonator}

\begin{figure}[!ht]
	\centering
	\includegraphics[width=\linewidth]{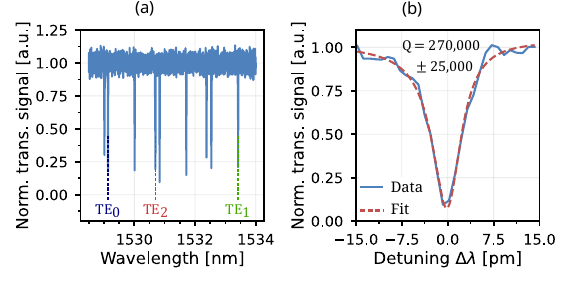}
	\caption{Transmission characterization of a multimode LN ring resonator with $w_r=5\,\mu\mathrm{m}$ and $\Delta_{\mathrm{wg}}=0.3\,\mu\mathrm{m}$. (a) Normalized transmission spectrum over two FSRs. (b) Fundamental TE$_0$ resonance near $\lambda=1531\,\mathrm{nm}$ with slightly asmymetric fano fit.}
	\label{fig:multimode_transmission}
\end{figure}

Fig.~\ref{fig:multimode_transmission} shows the transmission response of a multimode LN microring resonator with a larger rib width of $w_r=5\,\mu\mathrm{m}$ and a reduced negative lateral offset of $\Delta_{\mathrm{wg}}=-0.1$ µm, also chosen on the basis of its large transmission contrast. The normalized transmission spectrum recorded over two FSRs exhibits a significantly higher density of resonances, originating from the excitation of multiple transverse modes supported by the wider LN microring waveguide. The observed resonances are assigned to the TE$_0$, TE$_1$, and TE$_2$ modes based on their measured FSRs of $1.76$, $1.77$, and $1.76$~nm correspond to group indices of $n_g=2.118$, $2.108$, and $2.116$ for the TE$_0$, TE$_1$, and TE$_2$ modes, respectively.

The resonance associated with the fundamental TE$_0$ mode around $\lambda\approx1531\,\mathrm{nm}$ is analyzed in Fig.~\ref{fig:multimode_transmission} (b). A Fano fit of the experimental data, using the expression given in reference, yields a resonance linewidth of $\Gamma\approx5.7\,\mathrm{pm}$, corresponding to a loaded quality factor of $Q_\mathrm{L}=2.7\times10^{5}$, with a 95\% confidence interval of $\pm2.4\times10^{4}$. The extracted Fano asymmetry parameter indicates a nearly Lorentzian resonance profile. Considering the over-coupled regime predicted by simulations and a measured transmission minimum of $T_\mathrm{min}=0.117$, the intrinsic quality factor is estimated as $Q_\mathrm{i}\approx4.0\times10^{5}$ for the fundamental TE$_0$ mode.

\section{Conclusion and perspectives}

In conclusion, we demonstrate vertically-coupled whispering-gallery-mode resonators on LNOI using a wafer-scalable SU-8-based fabrication process, combining spin coating for thickness control and a controlled slow-ramp pre-exposure bake for planarization. The fabricated multimode and single-mode microring resonators exhibit intrinsic quality factors in agreement with predictions of $4\times10^{5}$ and $3.4\times10^{5}$, respectively. 

Further work will seek to capitalize on the advantageous features of the above-demonstrated integration scheme in particular for optical nonlinear conversion. Indeed, the demonstrated architecture is not only fully compatible with the geometry adjustments commonly used to induce dispersion management or phase-matching as required for nonlinear convertor demonstrators but also more amenable to implement the reduced gap coupling needed for multi-resonant devices (for instance, for SHG or Doubly-Resonant Optical Parametric Oscillators) that turns out to be otherwise technologically challenging to achieve. 
Additionally, the validated approach can also facilitate multilayer integration of cascaded ring resonators by vertically stacking resonant elements \cite{Bessette2013,Sumetsky2012}, providing new degrees of freedom for the design of complex linear or nonlinear photonic circuits.

\begin{backmatter}
\bmsection{Funding} ANR NanoFiLN (ANR-23-PEEL-004). 

\bmsection{Acknowledgment} This work was supported by LAAS-CNRS micro and nanotechnologies platform, a member
of the Renatech french national network. Part of the technological work was performed as part of OPALE, the joint laboratory between LAAS-CNRS and Essilor-Luxottica.

\smallskip

\bmsection{Disclosures} The authors declare no conflicts of interest.

\bmsection{Data availability} Data underlying the results presented in this paper are not publicly available at this time but may be obtained from the authors upon reasonable request.

\end{backmatter}

\bibliographystylefullrefs{opticajnlnt}
\bibliography{sample}

\begin{thebibliography}{10}
\newcommand{\enquote}[1]{``#1''}

\bibitem{Wang2019}
C.~Wang, M.~Zhang, M.~Yu, \emph{et~al.}, {\protect\JournalTitle{Nature
  Communications}}
  \href{https://www.nature.com/articles/s41467-019-08969-6}{\textbf{10}}
  (2019).

\bibitem{Moody2022}
G.~Moody, V.~J. Sorger, D.~J. Blumenthal, \emph{et~al.},
  {\protect\JournalTitle{Journal of Physics: Photonics}}
  \href{https://iopscience.iop.org/article/10.1088/2515-7647/ac1ef4}{\textbf{4}},
  012501 (2022).

\bibitem{Boes2023}
A.~Boes, L.~Chang, C.~Langrock, \emph{et~al.}, {\protect\JournalTitle{Science}}
  \href{https://www.science.org/doi/10.1126/science.abj4396}{\textbf{379}}
  (2023).

\bibitem{Younesi2025}
M.~Younesi, T.~Käsebier, I.~Elmanov, \emph{et~al.},
  {\protect\JournalTitle{Optical Materials Express}}
  \href{https://doi.org/10.1364/ome.540146}{\textbf{15}}, 299 (2025).

\bibitem{Nikogosyan2005}
D.~N. Nikogosyan, \emph{Nonlinear Optical Crystals: A Complete Survey}
  (Springer-Verlag, 2005).

\bibitem{Schiek2025}
R.~Schiek, A.~S. Solntsev, and D.~N. Neshev, {\protect\JournalTitle{Optical
  Materials Express}}
  \href{https://opg.optica.org/ome/fulltext.cfm?uri=ome-15-5-1100}{\textbf{15}},
  1100 (2025).

\bibitem{Cheng2023}
J.~Cheng, D.~Gao, J.~Dong, and X.~Zhang, {\protect\JournalTitle{Optics
  Express}} \href{https://doi.org/10.1364/oe.503988}{\textbf{31}}, 36736
  (2023).

\bibitem{Wu2024}
X.~Wu, Z.~Hao, L.~Zhang, \emph{et~al.}, {\protect\JournalTitle{Laser and
  Photonics Reviews}}
  \href{https://onlinelibrary.wiley.com/doi/10.1002/lpor.202300951}{\textbf{18}}
  (2024).

\bibitem{Chen2025}
T.~Chen, F.~Xue, R.~Hogan, \emph{et~al.},
  {\protect\JournalTitle{Nanophotonics}}
  \href{https://onlinelibrary.wiley.com/doi/10.1515/nanoph-2025-0211}{\textbf{14}},
  3023 (2025).

\bibitem{Koonath2004}
P.~Koonath, T.~Indukuri, and B.~Jalali, {\protect\JournalTitle{Applied Physics
  Letters}}
  \href{https://pubs.aip.org/apl/article/85/6/1018/745702/Vertically-coupled-micro-resonators-realized}{\textbf{85}},
  1018 (2004).

\bibitem{Calvez2015}
S.~Calvez, G.~Lafleur, A.~Larrue, \emph{et~al.}, {\protect\JournalTitle{IEEE
  Photonics Technology Letters}}
  \href{http://ieeexplore.ieee.org/lpdocs/epic03/wrapper.htm?arnumber=7050258}{\textbf{27}},
  982 (2015).

\bibitem{Feng2014}
J.~Feng and R.~Akimoto, {\protect\JournalTitle{IEEE Photonics Technology
  Letters}} \href{https://ieeexplore.ieee.org/document/6899635}{\textbf{26}},
  2391 (2014).

\bibitem{Ghulinyan2011}
M.~Ghulinyan, R.~Guider, G.~Pucker, and L.~Pavesi, {\protect\JournalTitle{IEEE
  Photonics Technology Letters}}
  \href{https://ieeexplore.ieee.org/document/5772911}{\textbf{23}}, 1166
  (2011).

\bibitem{Absil2001}
P.~Absil, J.~Hryniewicz, B.~Little, \emph{et~al.}, {\protect\JournalTitle{IEEE
  Photonics Technology Letters}}
  \href{https://ieeexplore.ieee.org/document/903217/}{\textbf{13}}, 49 (2001).

\bibitem{Kokubun2005}
Y.~Kokubun, Y.~Hatakeyama, M.~Ogata, \emph{et~al.}, {\protect\JournalTitle{IEEE
  Journal of Selected Topics in Quantum Electronics}}
  \href{https://ieeexplore.ieee.org/document/1395885/}{\textbf{11}}, 4 (2005).

\bibitem{zhang_MDPI2021}
J.~Zhang, R.~Wu, M.~Wang, \emph{et~al.}, {\protect\JournalTitle{Micromachines}}
  \href{https://www.mdpi.com/2072-666X/12/3/235}{\textbf{12}}, 235 (2021).

\bibitem{Niels2026}
M.~Niels, T.~Vandekerckhove, L.~De~Jaeger, \emph{et~al.},
  {\protect\JournalTitle{Nanophotonics}}
  \href{https://onlinelibrary.wiley.com/doi/10.1002/nap2.70042}{\textbf{15}}
  (2026).

\bibitem{Sumon2026}
M.~S.~I. Sumon, S.~Ghosh, A.~S. Vorobev, \emph{et~al.},
  {\protect\JournalTitle{IEEE Journal of Quantum Electronics}}
  \href{https://ieeexplore.ieee.org/document/11456956}{\textbf{62}}, 1 (2026).

\bibitem{Ruan2023}
Z.~Ruan, K.~Chen, Z.~Wang, \emph{et~al.}, {\protect\JournalTitle{Laser and
  Photonics Reviews}}
  \href{https://onlinelibrary.wiley.com/doi/10.1002/lpor.202200327}{\textbf{17}}
  (2023).

\bibitem{Selva2026}
C.~Selva, K.~Kinjalk, A.~Monmayrant, \emph{et~al.},
  {\protect\JournalTitle{Journal of the Optical Society of America B}}
  \textbf{43}, 2517 (2026).

\bibitem{Kaufmann2023}
F.~Kaufmann, G.~Finco, A.~Maeder, and R.~Grange,
  {\protect\JournalTitle{Nanophotonics}}
  \href{https://onlinelibrary.wiley.com/doi/10.1515/nanoph-2022-0676}{\textbf{12}},
  1601 (2023).

\bibitem{Gehring}
H.~Gehring, {\protect\JournalTitle{FEMWELL}} .

\bibitem{Optics}
A.~Optics, {\protect\JournalTitle{Lumerical}} .

\bibitem{Ghulinyan2013}
M.~Ghulinyan, F.~Ramiro-Manzano, N.~Prtljaga, \emph{et~al.},
  {\protect\JournalTitle{Physical Review Letters}}
  \href{https://journals.aps.org/prl/abstract/10.1103/PhysRevLett.110.163901}{\textbf{110}},
  163901 (2013).

\bibitem{Kinjalk2025}
K.~Kinjalk, C.~Selva, O.~Gauthier-Lafaye, \emph{et~al.},
  {\protect\JournalTitle{Optical Materials Express}}
  \href{https://doi.org/10.1364/ome.572573}{\textbf{15}}, 2826 (2025).

\bibitem{Ding14}
D.~Ding, M.~J.~A. de~Dood, J.~F. Bauters, \emph{et~al.},
  {\protect\JournalTitle{Opt. Express}}
  \href{https://opg.optica.org/oe/abstract.cfm?URI=oe-22-6-6778}{\textbf{22}},
  6778 (2014).

\bibitem{Bessette2013}
J.~T. Bessette and D.~Ahn, {\protect\JournalTitle{Optics Express}}
  \href{https://doi.org/10.1364/oe.21.013580}{\textbf{21}}, 13580 (2013).

\bibitem{Sumetsky2012}
M.~Sumetsky and Y.~Dulashko, {\protect\JournalTitle{Optics Express}}
  \href{https://doi.org/10.1364/oe.20.027896}{\textbf{20}}, 27896 (2012).

\end{thebibliography}

\bibliographyfullrefs{sample}

\end{document}